\documentclass[prd,aps,amsmath,amssymb,preprintnumbers,twocolumn,showpacs,floats]{revtex4}
\input{epsf}
\usepackage{epsfig}
\usepackage{graphics}
\usepackage{amsmath}
\usepackage{amsfonts}
\usepackage{amssymb}
\begin{document}
\title{Solitary waves  in Galilean covariant Fermi field theories with self-interaction}
\author{Fuad M. Saradzhev}
\affiliation{Centre for Science, Athabasca University, Athabasca,
Alberta, Canada\footnote{\tt E-mail address: fuads@athabascau.ca}}

\begin{abstract}

The generalized L\'{e}vy-Leblond equation for a $(3+1)$-dimensional self-interacting Fermi field is considered. Spin up solitary wave solutions with space oscillations in the $x^3$-coordinate are constructed. The solutions are shown to accumulate non-equal amounts of inertial and rest masses.

\end{abstract}


\maketitle


Solitary waves \cite{makh} in fermionic systems have been intensively studied for both the relativistic
and non-relativistic cases \cite{fink}-\cite{mert}. For non-relativistic systems, solitary waves have been constructed
as the solutions of the nonlinear Schrodinger (NLS) equation \cite{mert}. The NLS equation has been obtained
as the non-relativistic limit of the nonlinear Dirac equation and found important applications
in many areas, including plasma physics \cite{asan}, nonlinear optics \cite{islam} and molecular physics \cite{dav}.

In this paper, we follow a different approach to the study of solitary waves in non-relativistic
fermionic systems. Our starting point is the non-relativistic Dirac equation introduced by
L\'{e}vy-Leblond \cite{leblond}. We will refer to it as the L\'{e}vy-Leblond (LL) equation. It can be reduced to a
NLS type equation as well. However, it has a wider range of applications: it can be used in the regime
when the dynamics of non-relativistic fermions can not be obtained as the non-relativistic limit of a
relativistic dynamics. This happens when the inertial and rest masses of non-relativistic fermions are
not equal. The generalized version of the LL equation valid in this regime has been derived in \cite{fuad} within a $(4+1)$-dimensional Galilean covariant formulation of field theories \cite{khan}. Our aim is to construct solitary wave solutions for the generalized LL equation with the standard scalar-scalar self-interaction and to calculate the amounts of inertial and rest masses as well as energy and spin accumulated in these waves.


{\it Model}. We start with a five-dimensional Dirac field model given by the Lagrangian density:
\begin{equation}
 {\cal L}(x)= \overline{\Psi}(x)({\rm i} {\gamma}^{\mu}
{{\partial}_{\mu}}-k)\Psi(x) + g(\overline{\Psi}(x) {\Psi}(x))^2. \label{lagdirac}
\end{equation}
Herein ${\Psi}(x)$ is a  Dirac field defined on the five-dimensional manifold ${{\cal G}}_{\rm (4+1)}$ with the Galilean metric \cite{marc}
\begin{equation}
g_{\mu\nu}=\left( \begin{array}{ccc} -{\mathbf 1}_{3\times 3}
& 0 & 0 \\ 0 & 0 & 1 \\ 0 & 1 & 0\end{array} \right),\label{galileanmetric}
\end{equation}
 ${\mu},{\nu}=1,...,5$. The Dirac matrices $\gamma^{\mu}$ in the extended space-time are four-dimensional
\[
\gamma^a=
\left(\begin{array}{cc}
0 & {\rm i}\sigma^a\\
{\rm i}\sigma^a & 0
\end{array}\right),\;\;\;
\gamma^{4}=\frac{1}{\sqrt{2}}\left(
\begin{array}{cc}
1&1\\
-1&-1
\end{array}\right),
\]
\[
\gamma^5=\frac{1}{\sqrt{2}} \left(
\begin{array}{cc}
1&-1\\
1&-1
\end{array}\right),
\]
$a=1,2,3$, where ${\sigma}^1$,${\sigma}^2$, and ${\sigma}^3$ are the Pauli matrices,  and obey the anti-commutation relations:
\[
\left\{ \gamma^{\mu},\gamma^{\nu}\right\}=2g^{\mu \nu}.
\]
The adjoint field is defined as
\[
\overline{\Psi}(x)=\Psi^\dagger(x)\ \gamma^0,\]
where
\[
\gamma^0=\frac{1}{\sqrt{2}}\left(\gamma^4+\gamma^5\right)=
\left(
\begin{array}{cc}
1 & 0\\
0 & -1
\end{array}
\right).
\]
In Eq.(\ref{lagdirac}), $k$ is a momentum parameter related to the rest energy, while $g>0$ is the coupling constant.

The reduction to $(3+1)$-dimensions  performed by factoring the $x^5$-coordinate out of the original field
 ${\Psi}(x)$ as
\begin{equation}
\Psi (x) = e^{-{\rm i} m \bar{c} x^5}{\psi}_{+}({\bf x},t)
\label{factor}
\end{equation}
introduces the non-relativistic Fermi field ${\psi}_{+}({\bf x},t)$  with the positive inertial mass $m$ and  the rest energy $E_k \equiv k^2/(2m) = m_0 {\bar{c}}^2$. The parameters $m_0$ and $\bar{c}$ are the rest mass and velocity, respectively. The velocity parameter $\bar{c}$ can be identified with the speed of light if the inertial and rest masses are equal.

The Lagrangian density given by Eq.(\ref{lagdirac}) reduces to
\begin{eqnarray}
 {\cal L}_{+}({\bf x},t)
& = & \overline{\psi}_{+}({\bf x},t)( {\rm i} \gamma^{\bar{\mu}}
{\partial_{\bar{\mu}}}- kI_{+})\psi_{+}({\bf x},t) \nonumber\\
& + & g(\overline{\psi}_{+}({\bf x},t) {\psi}_{+}({\bf x},t))^2,
\label{newlagdirac}
\end{eqnarray}
where $\bar{\mu}$ runs from $1$ to $4$, and
\[
I_{+}=I - \frac{m\bar{c}}{k} {\gamma}^5,
\]
$I$ being the identity matrix. In the system of units where ${\hbar} = 1$, the combination of parameters $(gk/m)^{1/2}$ has the dimension of length. We will denote this combination as $L$.

From Eq.(\ref{newlagdirac}) we obtain the following equation of motion for the field ${\psi}_{+}({\bf x},t)$
\begin{equation}
\left({\rm i}
\gamma^{\bar{\mu}}\partial_{\bar{\mu}}-kI_{+}\right){\psi}_{+}({\bf x},t)= -2gA({\bf x},t) {\psi}_{+}({\bf x},t),
\label{nonreleq}
\end{equation}
where
\[
A({\bf x},t) \equiv \overline{\psi}_{+}({\bf x},t) {\psi}({\bf x},t).
\]
This is the generalized LL equation with self-interaction included. Representing the field ${\psi}_{+}({\bf x},t)$ as
\[
{\psi}_{+}({\bf x},t)=
\left(
\begin{array}{c}
{\psi}_{1,+}({\bf x},t) \\ {\psi}_{2,+}({\bf x},t)
\end{array}
\right),
\]
and introducing the linear combinations
\begin{eqnarray}
{\eta}_{1,+}({\bf x},t) & = & {\psi}_{1,+}({\bf x},t) + {\psi}_{2,+}({\bf x},t), \nonumber \\
{\eta}_{2,+}({\bf x},t) & = & {\psi}_{1,+}({\bf x},t) - {\psi}_{2,+}({\bf x},t), \nonumber
\end{eqnarray}
we bring  Eq.(\ref{nonreleq}) to the following form
\begin{eqnarray}
{\rm i} {\partial}_4 {\eta}_{1,+} +  p_{-} {\eta}_{2,+} & = &  -Ag\sqrt{2} {\eta}_{2,+}, \\
p_{+} {\eta}_{1,+} - m\bar{c} {\eta}_{2,+} & = &  Ag\sqrt{2} {\eta}_{1,+},
\label{system1}
\end{eqnarray}
where
\[
p_{\pm} \equiv \dfrac{1}{\sqrt{2}} ({\sigma}^a {\partial}_a \pm k),
\]
while $A({\bf x},t)$ becomes
\begin{equation}
A = \frac{1}{2} ({\eta}_{1,+}^{\dagger} {\eta}_{2,+} + {\eta}_{2,+}^{\dagger} {\eta}_{1,+}).
\label{exA}
\end{equation}
As in the case without self-interaction, the component ${\eta}_{2,+}$ is not dynamically independent. Its time evolution is completely determined by ${\eta}_{1,+}$.


{\it Solitary waves}. The generalized LL equation can be solved perturbatively in the weak-coupling regime. In this case, its solutions are expanded in powers of $g$. However, it is possible to solve the equation exactly for all values of $g$ if we restrict our consideration to the case of one-dimensional spin up solitary waves. Let us assume that both ${\eta}_{1,+}$ and ${\eta}_{2,+}$ are eigenvectors of ${\sigma}^3$ with the eigenvalue $(+1)$,
\[
{\sigma}^3 {\eta}_{1,+} = {\eta}_{1,+}, \qquad {\sigma}^3 {\eta}_{2,+} = {\eta}_{2,+},
\]
and that they both do not depend on $x^1$ and $x^2$ coordinates. Then
${\eta}_{1,+}$,${\eta}_{2,+}$ can be represented in the form
\[
{\eta}_{1,+}=
\left(
\begin{array}{c}
{\eta}_{1} \\ 0
\end{array}
\right),
\qquad
{\eta}_{2,+}=
\left(
\begin{array}{c}
{\eta}_{2} \\ 0
\end{array}
\right),
\]
where ${\eta}_1(x^3,t)$, ${\eta}_2(x^3,t)$ are single component fields.

Introducing the dimensionless variables
\[
{\tau} \equiv m\bar{c} x^4,  \qquad  {\xi}_{1,2} \equiv L{\Big(\frac{\mu}{E_k}\Big)}^{1/2}  e^{{\rm i}{\mu}^2{\tau}}  {\eta}_{1,2},
\]
\[
z \equiv m\bar{c} x^3,  \qquad {\cal A} \equiv L^2 \frac{\mu}{E_k}A,
\]
where ${\mu} \equiv \sqrt{m_{0}/m}$, we use Eq.(\ref{system1}) to express ${\xi}_2$ in terms of ${\xi}_1$ and ${\cal A}$ as
\begin{equation}
{\xi}_2 =\Big(\frac{1}{\sqrt{2}} \frac{\partial}{{\partial}z} + {\mu}\Big){\xi}_1 - {\cal A}{\xi}_1.
\label{elimination}
\end{equation}
Substituting next this expression into Eq.(\ref{exA}), this yields
\begin{equation}
{\cal A} = \frac{1}{2} \frac{1}{1+|{\xi}_1|^2}\Big(\frac{1}{\sqrt{2}} \frac{\partial}{{\partial}z} + 2{\mu}\Big)|{\xi}_1|^2.
\label{exA2}
\end{equation}
With equations (\ref{elimination}) and (\ref{exA2}), Eq.(6) takes the form of the non-linear Schr\"odinger equation
\begin{equation}
{\rm i}\frac{{\partial}{\xi}_1}{{\partial}{\tau}} + \frac{1}{2}  \frac{{\partial}^2{\xi}_1}{{\partial}z^2} + V{\xi}_1 = 0
\label{schrod}
\end{equation}
with the potential
\begin{equation}
V \equiv - \frac{1}{\sqrt{2}} \frac{{\partial}{\cal A}}{{\partial}z} + 2{\mu}{\cal A} - {\cal A}^2.
\label{potential}
\end{equation}

We look for a solution to this equation in the form of a wave propagating in the positive $z$-direction as
\begin{equation}
{\xi}_1(z,{\tau}) = D(z - {\nu}{\tau}) \exp\{\frac{\rm i}{2} ({\beta}^2 - {\nu}^2){\tau} + {\rm i} {\nu}z\},
\label{ansatz}
\end{equation}
where ${\beta}$ and ${\nu}$ are arbitrary parameters, while the function $D(z - {\nu}{\tau})$ is assumed to vanish as $|z| \to \infty$. Using it in Eq.(\ref{schrod}), this gives us
\begin{eqnarray}
\frac{d^2D}{d{\kappa}^2} - \frac{D}{1+D^2}{\Big(\frac{dD}{d{\kappa}}\Big)}^2 & + &\Big(2{\mu}^2 - {\beta}^2\Big) D(1+D^2) \nonumber \\
& - & 2 {\mu}^2 \frac{D}{1+D^2} = 0,
\label{eqD1}
\end{eqnarray}
where ${\kappa} \equiv z - {\nu}{\tau}$.

Integrating the last equation, we get
\begin{equation}
{\Big(\frac{dD}{d{\kappa}}\Big)}^2 = \Big({\beta}^2 - 2{\mu}^2\Big) D^2(1+D^2) - 2{\mu}^2 + C(1+D^2).
\label{eqD2}
\end{equation}
This simplifies to
\begin{equation}
\frac{dD}{d{\kappa}} = \pm \sqrt{{\beta}^2 D^2 +\Big({\beta}^2 - 2{\mu}^2\Big)D^4 }
\label{simple}
\end{equation}
if we take the constant of integration $C$ equal to $2{\mu}^2$. For
\begin{equation}
0<{\beta}^2 <2 {\mu}^2,
\label{condition}
\end{equation}
 the solution is
\begin{equation}
D = \frac{\beta}{\sqrt{2{\mu}^2 - {\beta}^2}} {\rm sech}\Big({\beta}{\kappa}\Big).
\label{Dsol}
\end{equation}

In terms of the original variables, the solitary wave solution is
\begin{eqnarray}
{\eta}_1 & = & \frac{1}{L} \sqrt{\frac{{E_k}(E_k - {\omega})}{{\mu}{\omega}}}  {\rm sech}\Big( \sqrt{2m(E_k - {\omega})}(x^3 - vt) \Big) \nonumber \\
& \times & \exp\left\{{\rm i}\Big({\omega} - \frac{1}{2}mv^2\Big)t + {\rm i}mv x^3 \right\},
\label{solwave}
\end{eqnarray}
where $v \equiv {\nu}\bar{c}$ is the speed of the solitary wave,
\[
{\omega} \equiv E_k \Big(1 - \frac{{\beta}^2}{2{\mu}^2} \Big)
\]
is the frequency of oscillations in time in the rest frame, while the condition given by Eq.(\ref{condition}) becomes
\[
0<{\omega}<E_k.
\]
The amplitude of the solitary wave increases with increasing of $E_k$ and/or decreasing of ${\omega}$.


{\it Mass, energy and spin}. Multiplying Eq.(\ref{nonreleq}) by $\overline{\psi}_{+}({\bf x},t)$ from the left and subtracting from it the adjoint equation multiplied by ${\psi}_{+}({\bf x},t)$ from
the right, we get
\begin{equation}
\frac{{\partial}{\rho}}{{\partial}t} + \frac{{\partial}j^a}{{\partial}x^a} = 0,
\label{coneq}
\end{equation}
where ${\rho} \equiv (1/{\bar{c}}) \overline{\psi}_{+} {\gamma}^4  {\psi}_{+}$ is the density of the non-relativistic Fermi field, and $j^a \equiv \overline{\psi}_{+} {\gamma}^a
{\psi}_{+}$ is its current density.

The total inertial mass of the system is defined as
\begin{equation}
M = m \int {\rho} d{\bf x} = \frac{m}{\bar{c}} \int  \overline{\psi}_{+} {\gamma}^4  {\psi}_{+} d{\bf x}.
\label{totmass}
\end{equation}
It is conserved in time for the solitary wave solution presented in Eq.(\ref{solwave}). The inertial mass accumulated in the solitary wave passing through the area ${\Omega} \equiv L^2$ in the $(x^1,x^2)$-plane is
\begin{equation}
M = m \frac{\sqrt{E_k(E_k - {\omega})}}{\omega}.
\label{wavemass}
\end{equation}

The Hamiltonian density corresponding to the Lagrangian density given by Eq.(\ref{newlagdirac}) is
\begin{eqnarray}
 {\cal H}_{+}({\bf x},t)
& = & - \overline{\psi}_{+}({\bf x},t)( {\rm i} \gamma^{a}
{\partial_{a}}- kI_{+})\psi_{+}({\bf x},t) \nonumber\\
& - & g(\overline{\psi}_{+}({\bf x},t) {\psi}_{+}({\bf x},t))^2.
\label{hamdensity}
\end{eqnarray}
This yields the following expression for the energy of the solitary wave passing through the same area ${\Omega}$:
\begin{eqnarray}
E & \equiv & \int {\cal H}_{+}({\bf x},t) d{\bf x} \nonumber\\
& = & E_0 +  \frac{1}{2} Mv^2,
\label{totenergy}
\end{eqnarray}
where
\begin{equation}
E_0 \equiv M_0 {\bar{c}}^2
\end{equation}
and
\begin{equation}
M_0 \equiv m_0 \Big[2 {\rm tanh}^{-1}\Big(\sqrt{1-\frac{\omega}{E_k}}\Big) - \sqrt{1-\frac{\omega}{E_k}} \Big]
\label{restmass}
\end{equation}
can be identified as the rest energy of the solitary wave and its rest mass, respectively.
\begin{figure}
\epsfig{file=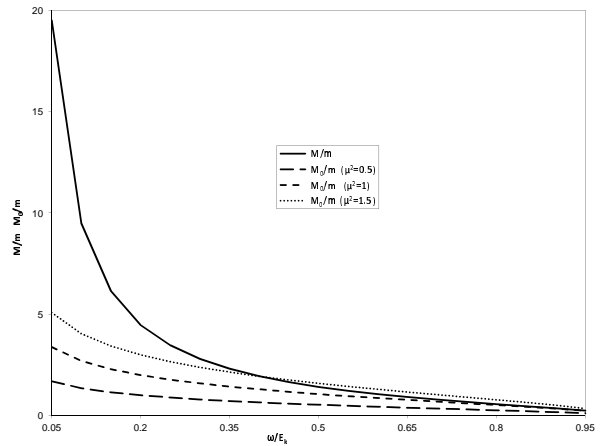,width=8.2cm,angle=-90}
\caption{The inertial and rest masses of the solitary wave as functions of ${\omega}/{E_k}$ at different values of ${\mu}^2$.}
\label{newgraph}
\end{figure}
%
%
%

As seen from FIG. \ref{newgraph}, both the inertal and rest masses decrease with increasing of ${\omega}/{E_k}$. However, the inertial mass decreases at a higher rate. The smaller ${\mu}^2$, the less the rest mass accumulated in the solitary wave. For, ${\mu}^2=1$, the inertial and rest masses practically coincide for values of ${\omega}/{E_k}$ close to $1$, while for ${\mu}^2>1$ this only happens at a single value of ${\omega}/{E_k}$.

According to the Vakhitov-Kolokolov criterion \cite{vako}, decreasing of the inertal mass with increasing of ${\omega}/{E_k}$ indicates that the solitary wave solution given by Eq.(\ref{solwave}) is stable for all values of ${\omega}$ in the interval $0<{\omega}<E_k$. This is in agreement with the statements about linear stability of solitary waves in the non-relativistic regime of the non-linear Dirac equation with the self-interaction
$g(\overline{\psi}_{+} {\psi}_{+})^2$ \cite{come}.

To find spin accumulated in the solitary wave, we use the Dirac spin operator defined as \cite{fuad}
\begin{equation}
{\cal S}^a = \frac{1}{2} \int d{\bf x} \frac{1}{\bar{c}}
\bar{\psi}_{+}({\bf x},t) {\gamma}^4 {\Sigma}^a {\psi}_{+}({\bf x},t),
\label{diracspin}
\end{equation}
where
\[
{\Sigma}^a \equiv
\left(\begin{array}{cc}
\sigma^a & 0\\
0 & \sigma^a
\end{array}\right).
\]
Since ${\Sigma}^3{\psi}_{+} = {\psi}_{+}$, the third component of the spin becomes
\begin{equation}
{\cal S}^3 = \frac{1}{2} \int {\rho} d{\bf x} = \frac{1}{2} \frac{\sqrt{E_k(E_k - {\omega})}}{\omega},
\label{spin3}
\end{equation}
while two other components vanish:
\[
{\cal S}^1 = {\cal S}^2 = 0.
\]
Instead of the Dirac spin operator, we could use the Galilean covariant spin operator as well \cite{fuad}. For one-dimensional solitary waves, these two operators coincide and have the same components.

The idea that the inertial and rest masses of a non-relativistic system are not necessarily the same has been known
for some time \cite{jack}. In our work, we have given an example of such systems.

The amounts of inertial and rest masses of the solitary waves are expressed in terms of the inertial and rest
masses of the original Fermi field, $m$ and $m_0$, respectively, and its rest energy $E_k$. These amounts are not equal to each other, even if $m_0$ is taken equal to $m$. For a given value of $E_k$, the solitary waves with a lower rest frame frequency of oscillations in time accumulate more inertial and rest masses and spin than the ones with higher frequencies.



\begin{thebibliography}{99}
\bibitem{makh} For a review, see, for instance, V.G.~Makhankov, Phys. Rep. {\bf 35}, 1 (1978); Y.~Kivshar, B.~Malomed,
Rev. Mod. Phys. {\bf 61}, 763 (1989); F.Kh.~Abdullaev, Phys. Rep. {\bf 179}, 1 (1989); M.~Remoissenet, {\it Waves
Called Solitons}, Springer-Verlag (1999); F.D.~Faddeev, L.~Takhtajan, {\it Hamiltonian Methods in the Theory
of Solitons}, Springer-Verlag (2007); M.J.~Ablowitz, {\it Nonlinear Dispersive Waves}, Cambridge Univ. Press
(2011).
\bibitem{fink} R.~Finkelstein, R.~Lelevier, M.~Ruderman, Phys. Rev. {\bf 83}, 326 (1951); R.J.~Finkelstein,
C.~Fronsdal, P.~Kaus, Phys. Rev. {\bf 103}, 1571 (1956); U.~Enz, Phys. Rev. {\bf 131}, 1392 (1963).
\bibitem{soler} M.~Soler, Phys. Rev. {\bf D1}, 2766 (1970); S.Y.~Lee, T.K.~Kuo, A.~Gavrielides, Phys. Rev. {\bf D12},
2249 (1975); A.~Alvarez, B.~Carreras, Phys. Lett. {\bf 86A}, 327 (1981); W.~Strauss, L.~Vazquez, Phys. Rev. {\bf D34},
641 (1986); Y.~Nogami, F.M.~Toyama, Phys. Rev. {\bf A45}, 5258 (1992).
\bibitem{vako} N.G.~Vakhitov, A.A.~Kolokolov, Radiophys. Quantum Electron. {\bf 16}, 783 (1973).
\bibitem{der} G.H.~Derrick, J. Math. Phys. {\bf 5}, 1252 (1964); I.L.~Bogolubsky, Phys. Lett. {\bf 73A}, 87 (1979);
A.~Alvarez, M.~Soler, Phys. Rev. Lett. {\bf 50}, 1230 (1983); J.~Xu, S.H.~Shao, H.Z.~Tang, J. Comput. Phys. {\bf 245},
131 (2013).
\bibitem{mert} F.G.~Mertens, N.~Quintero, A.R.~Bishop, Phys. Rev. {\bf E81}, 016608 (2010); F.~Cooper, A.~Khare,
B.~Mihaila, A.~Saxena, Phys. Rev. {\bf E82}, 036604 (2010); F.G.~Mertens, N.~Quintero, I.~Barashenkov,
A.R.~Bishop, Phys. Rev. {\bf E84}, 026614 (2011); F.~Cooper, A.~Khare, N.R.~Quintero, F.G.~Mertens, A.~Saxena,
Phys. Rev. {\bf E85}, 046607 (2012).
\bibitem{asan} N.~Asano, T.~Taniuti, N.~Yajima, J. Math. Phys. {\bf 10}, 2020 (1968); Y.H.~Ichikawa, T.~Imamura,
T.~Taniuti, J. Phys. Soc. Jpn. {\bf 33}, 189 (1972).
\bibitem{islam} R.I.~Chiao, E.~Garmire, C.H.~Townes, Phys. Rev. Lett. {\bf 13}, 479 (1964); P.L.~Kelley,
Phys. Rev. Lett. {\bf 15}, 1005 (1965); A.~Hasegawa, F.~Tappert, Appl. Phys. Lett. {\bf 23}, 142 (1973);
L.F.~Mollenauer, R.H.~Stolen, Opt. Lett. {\bf 9}, 13 (1984); H.A.~Haus, M.N.~Islam, IEEE J. Quantum Electron.
{\bf 21}, 1172 (1985).
\bibitem{dav} A.S.~Davydov, Phys. Scr. {\bf 20}, 387 (1979); {\it Solitons in Molecular Physics}, Kluwer (1991).
\bibitem{leblond} J.-M.~L\'{e}vy-Leblond, Commun. Math. Phys. {\bf 6}, 286 (1967).
\bibitem{fuad} F.M.~Saradzhev, Phys. Rev. {\bf D 84}, 125025 (2011).
\bibitem{khan} M.~de~Montigny, F.C.~Khanna, A.E.~Santana, E.S.~Santos, J.D.M.~Vianna, Ann. Phys. (N.Y.) {\bf 277}, 144 (1999); J. Phys. {\bf A33}, L273 (2000); M.~de~Montigny, F.C.~Khanna, A.E.~Santana, E.S.~Santos, J. Phys. {\bf A34}, 8901 (2001); M.~de~Montigny, F.C.~Khanna, A.E.~Santana, Int. J. Theor. Phys. {\bf 42}, 649 (2003); J. Phys. {\bf A36}, 2009 (2003); L.~Abreu, M.~de~Montigny, F.C.~Khanna, A.E.~Santana, Ann. Phys. (N.Y.) {\bf 308}, 244 (2003); E.S.~Santos, M.~de~Montigny, F.C.~Khanna, A.E.~Santana, J. Phys. {\bf A37}, 9771 (2004); E.S.~Santos, M.~de~Montigny, F.C.~Khanna, Ann. Phys. (N.Y.) {\bf 320}, 21 (2005).
\bibitem{marc} M.~de~Montigny, F.C.~Khanna, F.M.~Saradzhev, Ann. Phys. (N.Y.) {\bf 323}, 1191 (2008).
\bibitem{come} S.~Shao, N.R.~Quintero, F.G.~Mertens, F.~Cooper, A.~Khare, A.~Saxena, arXiv:1405.5547 [nlin.PS] and references therein.
\bibitem{jack} R.~Jackiw, V.P.~Nair, Phys. Lett. {\bf B480}, 237 (2000); {\bf B551}, 166 (2003); C.~Duval, P.A.~Horv\'{a}thy, Phys. Lett. {\bf B547}, 306 (2002); Erratum-ibid. {\bf B588}, 228 (2004).
\end{thebibliography}
\end{document}